\documentclass[onecolumn,
 amsmath,amssymb,
 aps,
 pra,
]{revtex4-2}

\usepackage{graphicx}% Include figure files
\usepackage{dcolumn}% Align table columns on decimal point
\usepackage{bm}% bold math
\usepackage{physics}
\usepackage{xcolor}
\usepackage[capitalize]{cleveref}
 \usepackage{appendix}
\begin{document}

\title{Spectral Characterization of Optical Aberrations  in  Fluidic Lenses }% Force line breaks with \\

\author{Graciana Puentes$^{1,2}$}
\author{Fernando Minotti$^{1,3}$}
% \email{gpuentes@df.uba.ar}
\affiliation{$^1$Universidad de Buenos Aires. Facultad de Ciencias Exactas y Naturales. Departamento de F\'{i}sica. Buenos Aires, Argentina \\
$^2$CONICET-Universidad de Buenos Aires. Instituto de F\'{i}sica de Buenos Aires (IFIBA). Buenos Aires, Argentina\\
$^3$CONICET-Universidad de Buenos Aires. Instituto de F\'{i}sica Interdisciplinaria y Aplicada (INFINA). Buenos Aires, Argentina.
}%

%\affiliation{3-Departamento de Física, Universidade %Federal de Minas Gerais, Belo Horizonte, MG 31270-901, %Brazil
%}%

\begin{abstract}
We report an \color{black} extensive \color{black} numerical study and supporting experimental results on the spectral characterization of optical aberrations in macroscopic fluidic lenses with tunable focal distance and aperture shape. Using a Shack-Hartmann wave-front sensor we experimentally 
reconstruct the near-field wave-front transmitted by the fluidic lenses, and we characterize the chromatic aberrations in terms of Zernike polynomials in the visible range. Moreover, we further classify the spectral response of the lenses
 using clustering techniques, in addition to correlation and convolution measurements. Experimental results are in agreement with our theoretical model of the non-linear deformation of thin elastic membranes. 
\end{abstract}

\maketitle

%\tableofcontents
\section{Introduction}

One of the most common ocular disorders worldwide, and the main cause of visual impairment in children, is myopia. The elongation of the axial length in the eyes which characterizes medium and high levels of myopia can increase the risk of severe ocular pathologies, potentially leading to irreversible blindness. Most traditional adaptive eye-wear based on fluidic lenses aim to correct refractive errors requiring medium dioptric power, such as mild myopia, hyperopia, an other focus errors \cite{WO2006,Mastrangelo}. On the other hand, refractive errors other than focus, including coma, astigmatism and higher order aberrations are usually treated via astigmatic corrections \cite{Gross, Keating, Tasman}, which are more difficult to achieve with standard fluidic lenses. Moreover, in most cases compounded errors are present, most commonly presbyopia with focus defects, requiring multi-focal lenses whose limited accommodation distance and highly restricted field-of-view can lead to high loss of visual capacity \cite{Johnson}. Finally, those patients with severe visual impairment due to glaucoma or other visual traumas, require large dioptric power corrections, necessitating thick organic lenses, which are prone to high-order aberrations, in addition to being significantly unattractive and unpractical.
  In a previous publication \cite{OSACPuentes}, we presented the first macroscopic fluidic lens eye-wear prototype with high dioptric power (+25D to +100D range) with optical aberrations below a fraction of the wavelength, 
  which can adaptively restore accommodation distance within several centimeters, thus enabling access to the entire field-of-view. The lens is made of an elastic polymer of the PDMS type which can adaptively modify its optical power according to the 
  fluidic volume mechanically pumped in. Such liquid lens exhibits a large dynamic range, and its focusing properties are polarization independent  \cite{Puentes}. Additionally, we demonstrated that by tuning the lens aperture it is possible to address different optical aberrations, thus providing an additional degree of freedom for the lens design. Our design is attractive for adaptive eye-wear, in addition to cellular phone, camera, optical zooms, or other machine vision applications where large magnification  can be required \cite{PuentesPatent}. \\

%Adaptive fluidic lenses are customarily classified in two main categories, according to their tuning mechanism. The first category is the electro-wetting lens whose focal length can be tuned continuously by applying a controlled external voltage \cite{Vallet, Krupenkin, Kuiper}. 
%While the advantage of electro-wetting lenses is that they can provide for large focusing power with no mechanical moving parts, the disadvantage is the high driving voltage required, in addition to limited stability and liquid evaporation due to heating. The second type is the mechanical lens, whose focal length is controlled by pumping fluid in the elastic lens chamber, thus introducing lens curvature changes \cite{Knollman, Sugiura, Zhang, Chronis, Moran, Ren}. The adaptive lenses reported here belong to the second category. For this demonstration of principle, we pump the fluid in by using a syringe pump. We note that  the mechanical pump itself is not the focus of this Report. \\

In this paper, we present an  \color{black} extensive \color{black} numerical and experimental spectral study of optical aberrations in macroscopic fluidic lenses with high dioptric power, tunable focal distance and aperture shape \cite{OSACPuentes}, based on an empirical characterization of the refractive index of thin elastic membranes, such as PDMS, according to the Sellmeier model \cite{Schneider}. Using a Shack-Hartmann wave-front sensor we experimentally 
reconstruct the near-field wave-front transmitted by such fluidic lenses, and we characterize the chromatic aberrations in terms of Zernike polynomials \color{black} over the visible wavelength range ($\lambda=400-650 $ nm), \color{black} using a programmable LED source. Moreover, we further classify the spectral response of the lenses
 using clustering techniques, in addition to correlation and convolution measurements. Experimental results are in  agreement with our theoretical model of the non-linear elastic membrane deformation.

\section{Theoretical model}

\color{black}\subsection{Inclusion of gravity effects}

We briefly recall the model used in \cite{OSACPuentes} to simulate the fluid lens surface shape without considering gravity
effects.\color{black} The equations used are those derived by Berger \cite{berger} to determine the nonlinear, large deformation of thin isotropic elastic plates: 
\begin{subequations}
\label{berger}
\begin{eqnarray}
\nabla ^{4}w-\alpha ^{2}\nabla ^{2}w &=&\frac{q}{D},  \label{b1} \\
\frac{\partial u}{\partial x}+\frac{\partial v}{\partial y}+\frac{1}{2}
\left( \frac{\partial w}{\partial x}\right) ^{2}+\frac{1}{2}\left( \frac{
\partial w}{\partial y}\right) ^{2} &=&\frac{\alpha ^{2}h^{2}}{12}.
\label{b2}
\end{eqnarray}
In these equations $w\left( x,y\right) $ is the local $z$-displacement of
the membrane, with non-deformed state assumed\ to correspond to the $z=0$
plane, $u\left( x,y\right) $ and $v\left( x,y\right) $ are the local $x$ and 
$y$ displacements, $D$ the membrane bending rigidity, and $h$ its thickness.
The magnitude $q\left( x,y\right) $ corresponds to the applied $z$-load, and 
$\alpha $ is a constant to be determined from the same equations by imposing appropriate boundary conditions.

For the case of uniform load (constant $q$) and elliptic aperture,
analytical solutions of the system (\ref{berger}) were obtained by the
method of constant deflection contour lines derived by Mazumdar \cite{mazumdar1}. If the aperture in the plane $z=0$ is an ellipse of $x$, $y$
semiaxes $a$ and $b$, respectively, the $z$-displacement of the membrane is given by 
\end{subequations}
\begin{equation}
w\left( \varsigma \right) =\frac{\Delta V}{\pi ab}\frac{2\gamma \left[
\gamma \left( 1-\varsigma ^{2}\right) I_{1}\left( 2\gamma \right)
+I_{0}\left( 2\gamma \varsigma \right) -I_{0}\left( 2\gamma \right) \right] 
}{\left( \gamma ^{2}+2\right) \,I_{1}\left( 2\gamma \right) -2\gamma
I_{0}\left( 2\gamma \right) }.  \label{w_zeta_DV}
\end{equation}
where $\Delta V$ is the volume of the liquid, the variable $\varsigma $ is defined as 
\begin{equation}
\varsigma ^{2}=x^{2}/a^{2}+y^{2}/b^{2},  \label{psi_ellipse}
\end{equation}
and the constant $\gamma $ is related to $\Delta V$ by 
\begin{equation}
\Delta V=\pi abh\frac{\sqrt{3a^{4}+2a^{2}b^{2}+3b^{4}}}{a^{2}+b^{2}}G\left(
\gamma \right) ,  \label{gamma_DV}
\end{equation}
where 
\begin{equation}
G\left( \gamma \right) =\frac{\left( \gamma ^{2}+2\right) \,I_{1}\left(
2\gamma \right) -2\gamma I_{0}\left( 2\gamma \right) }{\sqrt{24\gamma }\sqrt{
3\gamma \left[ I_{1}\left( 2\gamma \right) \right] ^{2}-2I_{2}\left( 2\gamma
\right) \left[ \gamma I_{0}\left( 2\gamma \right) +2I_{1}\left( 2\gamma
\right) \right] }}.  \label{G_gamma}
\end{equation}

\begin{figure}[h!]
\centering
\includegraphics[width=0.5\linewidth]{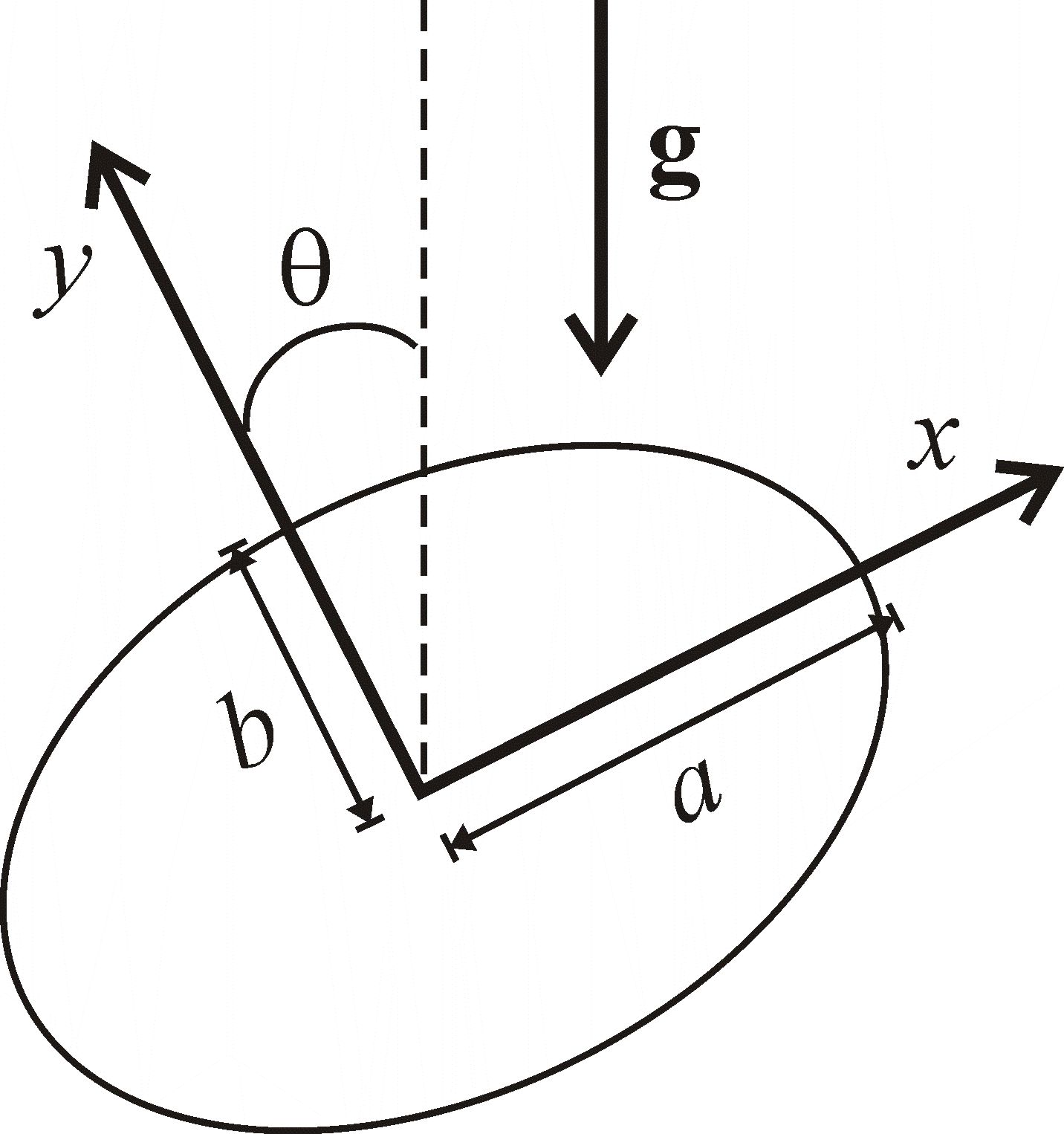}
\caption{Notation used in the model to include the effect of gravity on the lens surface shape.}
\label{fig:vertlens}
\end{figure}

\color{black}We consider now the effect of gravity when the ($x,y$) plane of the lens aperture is vertical. In Berger equations (\ref{berger}) the load $q$ has now the expression \color{black}
\begin{equation*}
q=q_{0}-\rho g\left( x\sin \theta +y\cos \theta \right) ,
\end{equation*}
with $q_{0}$ the load at the lens center ($x=y=0$), $\rho $ the mass density
of the filling fluid, $g$ the acceleration of gravity, and $\theta $ the
angle between the vertical direction and the $y$ axis \color{black}(see figure \ref{fig:vertlens})\color{black}.

In the case that $\rho gL/q_{0}\ll 1$, with $L$ a characteristic length of
the lens pupil, we can treat gravity effects as a perturbation to the case
with uniform load $q_{0}$, previously obtained, and write 
\begin{eqnarray*}
w &=&w_{0}+w_{1}, \\
\alpha &=&\alpha _{0}+\alpha _{1},
\end{eqnarray*}
with $w_{0}$ and $\alpha _{0}$ corresponding to the solution of the case
with $q=q_{0}$. Linearization of Eq. (\ref{b1}) in the perturbations $w_{1}$ and $\alpha _{1}$ yields 
\begin{equation}
\nabla ^{4}w_{1}-\alpha _{0}^{2}\nabla ^{2}w_{1}-2\alpha _{0}\alpha
_{1}\nabla ^{2}w_{0}=-\frac{\rho g}{D}\left( x\sin \theta +y\cos \theta
\right) ,  \label{w_linear}
\end{equation}

For a clamped membrane with no pre-stretching $u=v=0$ at $\psi =0$, and so
the integral of the linearized version of Eq. (\ref{b2}) over the area $S_{0}
$ of the lens aperture gives 
\begin{equation}
\int\limits_{S_{0}}\left( \frac{\partial w_{0}}{\partial x}\frac{\partial
w_{1}}{\partial x}+\frac{\partial w_{0}}{\partial y}\frac{\partial w_{1}}{
\partial y}\right) dxdy=\frac{\alpha _{0}\alpha _{1}h^{2}}{6}S_{0}.
\label{apha_linear}
\end{equation}

We now consider the solution to Eq. (\ref{w_linear}) in the usual case in
which the membrane forces dominate: $\alpha _{0}^{2}L^{2}\gg 1$. In this
case, except extremely close to the membrane border, one has $\left\vert
\nabla ^{4}w_{1}\right\vert \ll \left\vert \alpha _{0}^{2}\nabla
^{2}w_{1}\right\vert $, and so it is easy to check that the solution to Eq. (
\ref{w_linear}) satisfying the boundary condition $w_{1}=0$ at the border of
the membrane is given by 
\begin{equation}
w_{1}=-\frac{2\alpha _{1}}{\alpha _{0}}w_{0}-\frac{\rho ga^{2}b^{2}}{
2D\alpha _{0}^{2}}\left( 1-\frac{x^{2}}{a^{2}}-\frac{y^{2}}{b^{2}}\right)
\left( \frac{x\sin \theta }{a^{2}+3b^{2}}+\frac{y\cos \theta }{3a^{2}+b^{2}}
\right) .  \label{w1_solution}
\end{equation}

Using expression (\ref{w1_solution}) in Eq. (\ref{apha_linear}) one readily
obtains $\alpha _{1}=0$, so that the complete solution with the inclusion of
gravity effects is 
\begin{equation}
w=w_{0}-\frac{\rho g}{2D\gamma ^{2}}\frac{a^{4}b^{4}\left(
a^{2}+b^{2}\right) }{3a^{2}+2a^{2}b^{2}+3b^{4}}\left( 1-\frac{x^{2}}{a^{2}}-
\frac{y^{2}}{b^{2}}\right) \left( \frac{x\sin \theta }{a^{2}+3b^{2}}+\frac{
y\cos \theta }{3a^{2}+b^{2}}\right) ,  \label{w_full}
\end{equation}
where $w_{0}$ is the solution previously obtained for uniform load, Eq. (\ref
{w_zeta_DV}), and the constant $\gamma $ is the one determined in that
solution using Eq. (\ref{gamma_DV}).

\color{black}\subsection{Determination of the aberrations of the fluid lens}

\begin{figure}[h!]
\centering
\includegraphics[width=0.4\linewidth]{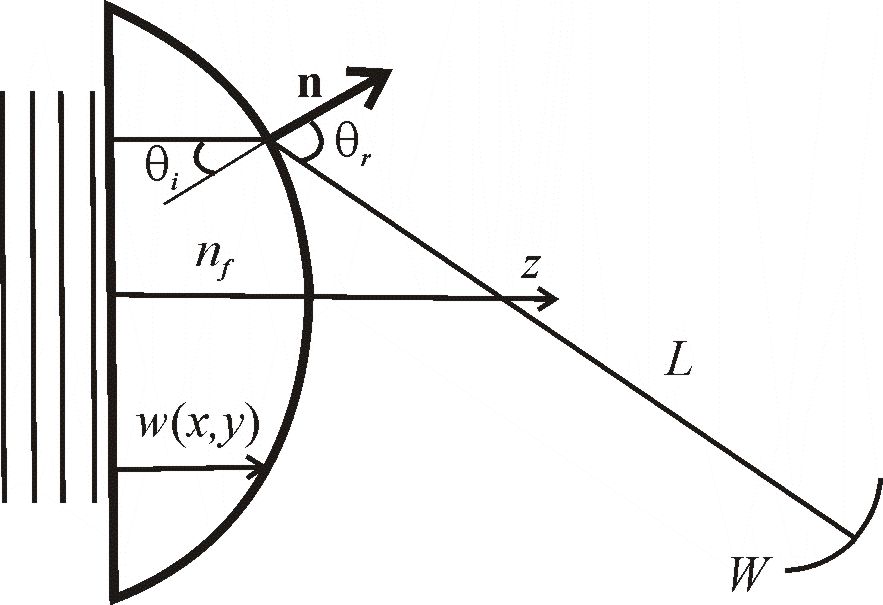}
\caption{Notation used in the model for the ray tracing from a plane wavefront incident on the plane surface of the lens to the refracted wavefront $W$}
\label{fig:aberration}
\end{figure}

In order to determine the aberrations of a fluid lens with one plane surface we
consider a plane wave front with normal incidence on the plane side of the
membrane, taken at $z=0$ (see figure \ref{fig:aberration}). The corresponding rays,
parallel to the $z$ -axis, of unit vector $\mathbf{e}_{z}$, are then
refracted according to Snell law when they cross the membrane curved surface
at $z=w\left( x,y\right) $. \color{black}The external normal unit vector at that surface
is given by (in Cartesian components) 
\begin{equation}
\mathbf{n}=\frac{\left( -w_{x},-w_{y,}1\right) }{\sqrt{1+w_{x}^{2}+w_{y}^{2}}},
\end{equation}
where the subscripts $x$, $y$ indicate derivatives with respect to the corresponding coordinate.

The angle $\theta _{i}$ of the rays incident from inside the lens, relative
to the external normal direction at the corresponding point of the membrane curved
surface, is thus given by 
\begin{equation}
\cos \theta _{i}=\mathbf{n}\cdot \mathbf{e}_{z}=\frac{1}{\sqrt{
1+w_{x}^{2}+w_{y}^{2}}}.
\end{equation}

Snell law then determines the angle of the refracted ray emerging from the
lens, also relative to the normal direction, as 
\begin{equation}
\sin \theta _{r}=n_{f}\frac{\sqrt{w_{x}^{2}+w_{y}^{2}}}{\sqrt{
1+w_{x}^{2}+w_{y}^{2}}},
\end{equation}
where $n_{f}$ is the index of refraction of the filling fluid, relative to
that of air.

The refracted ray is contained in the plane determined by the normal unit
vector $\mathbf{n}$ and the unit vector tangent to the surface 
\begin{equation}
\mathbf{t}=\frac{\mathbf{e}_{z}-\left( \mathbf{n}\cdot \mathbf{e}_{z}\right) 
\mathbf{n}}{\left\vert \mathbf{e}_{z}-\left( \mathbf{n}\cdot \mathbf{e}
_{z}\right) \mathbf{n}\right\vert }=\frac{\left(
w_{x},w_{y,}w_{x}^{2}+w_{y}^{2}\right) }{\sqrt{\left(
1+w_{x}^{2}+w_{y}^{2}\right) \left( w_{x}^{2}+w_{y}^{2}\right) }},
\end{equation}
so that the ray direction is given by the unit vector 
\begin{equation}
\mathbf{k}_{r}=\mathbf{n}\cos \theta _{r}+\mathbf{t}\sin \theta _{r}.
\end{equation}
The explicit expressions of the Cartesian components of $\mathbf{k}_{r}$\
are 
\begin{subequations}
\begin{eqnarray}
k_{rx,y} &=&\frac{n_{f}-\sqrt{1+\left( 1-n_{f}^{2}\right) \left(
w_{x}^{2}+w_{y}^{2}\right) }}{1+w_{x}^{2}+w_{y}^{2}}w_{x,y}, \\
k_{rz} &=&\frac{n_{f}\left( w_{x}^{2}+w_{y}^{2}\right) +\sqrt{1+\left(
1-n_{f}^{2}\right) \left( w_{x}^{2}+w_{y}^{2}\right) }}{1+w_{x}^{2}+w_{y}^{2}},
\end{eqnarray}
which are functions of the point ($x$, $y$) in the plane $z=0$, at which the ray originated.

In this way, a generic ray starting at the point ($x$, $y$) in the plane $
z=0 $ inside the lens is refracted at the point $\mathbf{X}_{0}=\left(
x,y,w\left( x,y\right) \right) $ on the membrane surface, and after
traversing in air a distance $L$ reaches the point $\mathbf{X}_{L}=\mathbf{X}
_{0}+\mathbf{k}_{r}L$, so that (from now on we do not write the explicit
dependence on ($x$, $y$) of $w$ and of $\mathbf{k}_{r}$) 
\end{subequations}
\begin{equation}
z_{L}=w+k_{rz}L,
\end{equation}
and 
\begin{subequations}
\begin{eqnarray}
x_{L} &=&x+\frac{k_{rx}}{k_{rz}}\left( z_{L}-w\right) , \\
y_{L} &=&y+\frac{k_{ry}}{k_{rz}}\left( z_{L}-w\right) .
\end{eqnarray}

The phase at ($x_{L}$, $y_{L}$, $z_{L}$) is thus 
\end{subequations}
\begin{equation}
\phi _{L}=\phi _{0}+\frac{2\pi }{\lambda }\left( n_{f}w+\frac{z_{L}-w}{k_{rz}
}\right) ,  \label{phase}
\end{equation}
where $\phi _{0}$ is the phase of the front at $z=0$, and $\lambda $ is the
wavelength in air.

From (\ref{phase}) we can determine the $z_{W}$ position of a wavefront of
given phase $\phi _{W}$ as
\begin{equation}
z_{W}=w\left( 1-n_{f}k_{rz}\right) +k_{rz}\frac{\lambda }{2\pi }\left( \phi
_{W}-\phi _{0}\right) ,  \label{zW}
\end{equation}
to which correspond the ($x_{W}$, $y_{W}$) coordinates 
\begin{subequations}
\begin{eqnarray}
x_{W} &=&x+k_{rx}\left[ \frac{\lambda }{2\pi }\left( \phi _{W}-\phi
_{0}\right) -n_{f}w\right] , \\
y_{W} &=&y+k_{ry}\left[ \frac{\lambda }{2\pi }\left( \phi _{W}-\phi
_{0}\right) -n_{f}w\right] .
\end{eqnarray}

These two relations can in principle be solved to give 
\end{subequations}
\begin{subequations}
\begin{eqnarray}
x &=&x\left( x_{W},y_{W}\right) , \\
y &=&y\left( x_{W},y_{W}\right) ,
\end{eqnarray}
which if replaced in (\ref{zW}) give the wavefront geometry: $
z_{W}=z_{W}\left( x_{W},y_{W}\right) $.

If this wavefront is analyzed at a position $z_{A}$ we can, without loss of
generality, take this position as that of the image of the origin, $x=y=0$: $
z_{A}=w_{0}\left( 1-n_{f}k_{rz0}\right) +k_{rz0}\frac{\lambda }{2\pi }\left(
\phi _{W}-\phi _{0}\right) $, where $w_{0}=w\left( 0,0\right) $ and $
k_{rz0}=k_{rz}\left( 0,0\right) $, so that 
\end{subequations}
\begin{equation}
\frac{\lambda }{2\pi }\left( \phi _{W}-\phi _{0}\right) =\frac{
z_{A}-w_{0}\left( 1-n_{f}k_{rz0}\right) }{k_{rz0}},
\end{equation}
and analyze the deviation from a plane front: $\Delta z_{W}=z_{W}-z_{A}$,
which is conveniently written as
\begin{equation}
\Delta z_{W}=w\left( 1-n_{f}k_{rz}\right) -\frac{k_{rz}}{k_{rz0}}w_{0}\left(
1-n_{f}k_{rz0}\right) +\left( \frac{k_{rz}}{k_{rz0}}-1\right) z_{A}.
\label{Dzw}
\end{equation}
The corresponding ($x_{W}$, $y_{W}$) coordinates are written as 
\begin{subequations}
\label{xyw}
\begin{eqnarray}
x_{W} &=&x+\frac{k_{rx}}{k_{rz0}}\left[ z_{A}-w_{0}-n_{f}k_{rz0}\left(w-w_{0}\right) \right] , \\
y_{W} &=&y+\frac{k_{ry}}{k_{rz0}}\left[ z_{A}-w_{0}-n_{f}k_{rz0}\left(w-w_{0}\right) \right].
\end{eqnarray}

\color{black}In this way, Eqs. (\ref{Dzw}) and (\ref{xyw}) give the wavefront geometry,
analyzed at $z=z_{A}$, parameterized in terms of the ($x$, $y$) coordinates on the plane at $z=0$.

We further model the wave-front analyzer at $z=z_{A}$ as having a circular
aperture of radius $r_{A}$ so that the section of the wave-front $\Delta
z_{W}\left( x_{W},y_{W}\right) $ to be studied is expressed as $\Delta
z_{W}\left( r_{A}\xi \cos \phi ,r_{A}\xi \sin \phi \right) $, with $0\leq
\xi \leq 1$, and decomposed in Zernike polynomials in polar coordinates $
Z_{n}\left( \xi ,\phi \right) $,

\end{subequations}
\begin{equation}
\Delta z_{W}=\sum_{n}a_{n}Z_{n}\left( \xi ,\phi \right) ,
\end{equation}
\color{black}with
\begin{equation}
a_{n}=\frac{1}{\pi }\int_{0}^{2\pi }\int_{0}^{1}\Delta z_{W}\left( r_{A}\xi
\cos \phi ,r_{A}\xi \sin \phi \right) Z_{n}\left( \xi ,\phi \right)\xi d\xi
d\phi .
\end{equation}

We have followed the standard OSA/ANSI indexing and normalization scheme, used in the Shack-Hartmann wave-front sensor, for which the first 15-term orthonormal Zernike circle polynomials are:
\begin{eqnarray*}
Z_{0} &=&1, \\
Z_{1} &=&2\xi \sin \phi , \\
Z_{2} &=&2\xi \cos \phi , \\
Z_{3} &=&\sqrt{6}\xi ^{2}\sin 2\phi , \\
Z_{4} &=&\sqrt{3}\left( 2\xi ^{2}-1\right) , \\
Z_{5} &=&\sqrt{6}\xi ^{2}\cos 2\phi , \\
Z_{6} &=&\sqrt{8}\xi ^{3}\sin 3\phi , \\
Z_{7} &=&\sqrt{8}\left( 3\xi ^{3}-2\xi \right) \sin \phi , \\
Z_{8} &=&\sqrt{8}\left( 3\xi ^{3}-2\xi \right) \cos \phi , \\
Z_{9} &=&\sqrt{8}\xi ^{3}\cos 3\phi , \\
Z_{10} &=&\sqrt{10}\xi ^{4}\sin 4\phi , \\
Z_{11} &=&\sqrt{10}\left( 4\xi ^{4}-3\xi ^{2}\right) \sin 2\phi , \\
Z_{12} &=&\sqrt{5}\left( 6\xi ^{4}-6\xi ^{2}+1\right) , \\
Z_{13} &=&\sqrt{10}\left( 4\xi ^{4}-3\xi ^{2}\right) \cos 2\phi , \\
Z_{14} &=&\sqrt{10}\xi ^{4}\cos 4\phi ,
\end{eqnarray*}
\color{black}

\begin{figure}[t!]
\centering
\includegraphics[width=1\linewidth]{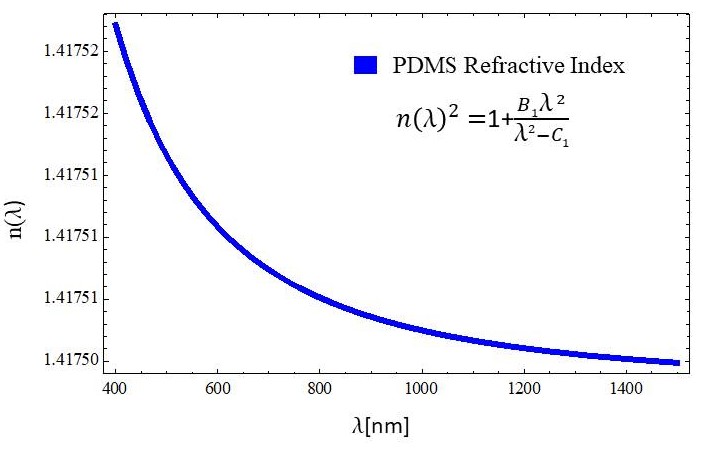}
\caption{PDMS refractive index ($n(\lambda)$) vs. wavelength ($\lambda$). The refractive index is obtained by  experimentally determining the Sellmeier coefficients ($B_1, C_1$), resulting in $n(\lambda)^2=1 + \frac{B_1 \lambda^2}{\lambda^2-C_1}$, with $B_1=1.0093$ and $C_1[nm^2]=13.185$ \cite{Schneider}.}
\label{fig:index}
\end{figure}

\begin{figure}[b!]
\centering
\includegraphics[width=1\linewidth]{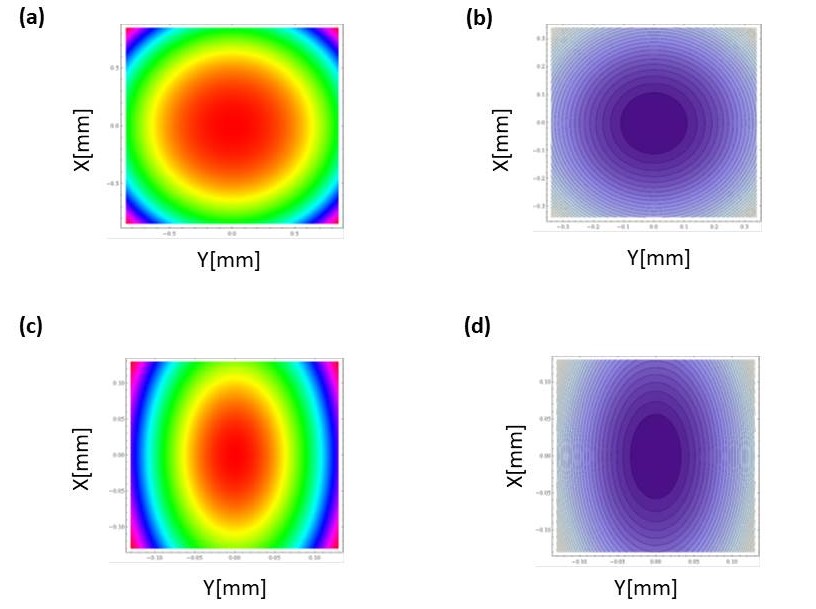}
\caption{(a) and (c) Density plots of acquired phase $P(x,y)$ upon propagation over a distance $z_{d}=3.5$ cm,  for $\lambda=600$ nm. (b) and (d) corresponding contour plots. Top row: circular aperture with horizontal  and vertical axes $a=b=1.7$ cm. Bottom row: elliptic aperture with  horizontal and vertical axes $a=1.3$ cm and $b=1.7$ cm, respectively.}
\label{fig:phase}
\end{figure}

%\begin{figure}[t!]
%\hspace{-0.6cm}
%\centering
%\includegraphics[width=1.0\linewidth]{Diapositiva3new.jpg}
%\caption{Numerical simulations of optical aberrations in terms of Zernike polynomials over the entire visible range $\lambda=400-1500$ nm, for fluidic lenses with circular aperture (axes $a=b=1.7$ cm). (a), (b), (c) and (d), correspond to unitless Zernike coefficients for polynomials of orders $P_0, P_2, P_3, P_8$, respectively. Zernike coefficients characterizing the remaining polynomials are negligible ($<< 10^{-17}$). Further details are in the text.  }
%\label{fig:false-color}
%\end{figure}

%\begin{figure}[h!]
%\hspace{-0.6cm}
%\centering
%\includegraphics[width=1.0\linewidth]{Diapositiva4new.jpg}
%\caption{Numerical simulations of optical aberrations in terms of Zernike polynomials over the entire visible range $\lambda=400-1500$ nm, for fluidic lenses with elliptic apertures characterized by ellipse axes ($a=1.5$ cm, $b=1.7$ cm, red curve) and 
%($a=1.3$ cm, $b=1.7$ cm, blue curve). (a), (b), (c) and (d), correspond to unitless Zernike coefficients for polynomials of orders $P_0, P_3, P_4, P_8$, respectively. Zernike coefficients characterizing the remaining polynomials are negligible ($<< 10^{-17}$). Further details are in the text.}
%\label{fig:false-color}
%\end{figure}

\section{Spectral response} 

In order to characterize the spectral response of the  polydimethylsiloxane (PDMS type) elastic membrane used to fabricate the fluidic lenses, we incorporate an empirical expression for the refractive index of PDMS Sylgard 184, as reported in \cite{Schneider}. The refractive index $n(\lambda)$ is decreasing for increasing wavelength $\lambda$, 
which is typical of glass and polymeric materials. For the approximation of the dispersion across the entire visible light spectrum, the Sellmeier dispersion model is used, which describes the empirical relation between the refractive index $n(\lambda)$ and the wavelength $\lambda$, given by:

\begin{equation}
n(\lambda)^2=1 + \frac{B_1 \lambda^2}{\lambda^2-C_1}+\frac{B_2 \lambda^2}{\lambda^2-C_2}+\frac{B_3 \lambda^2}{\lambda^2-C_3},
\end{equation}

where $B_1, B_2, B_3, C_1, C_2, C_3$ are the experimentally determined Sellmeier coefficients. As reported in Ref. \cite{Schneider}, $B_1=1.0093$ and $C_1[nm^2]=13.185$. \color{black} Due to the limited number of measurement points (three wavelengths with eight measurements each) the second and third Sellmeier coefficients are set to zero \cite{Schneider}. \color{black} A plot of the refractive index vs. wavelength within the range 400 nm to 1500 nm is presented in Figure \ref{fig:index}.\\

According to the theoretical model, the spectral response of the phase $P(x,y,\lambda)$ acquired by the beam upon propagation over a distance $z_d$ can be expressed as:

\begin{equation}
P(x,y,\lambda)=\frac{2 \pi}{\lambda}[ n(\lambda)* w + (z_{d}-w)*k_{z}(\lambda)],
\end{equation}

where $w$ is the local displacement in the $z$-direction, and $k_{z}$ is given by:
\begin{equation}
k_{z}(\lambda)= \frac{1+ (w_{x}^2+ w_{y}^2)}{n(\lambda) (w_{x}^2+ w_{y}^2)+ \sqrt{1+ (1-n(\lambda)^2) (w_{x}^2+ w_{y}^2)}}.
\end{equation}

We performed numerical simulations of the phase acquired by the beam upon propagation over a distance $z_{d}=3.5$ cm. Numerical simulations are displayed in Figure \ref{fig:phase}. Density plots of $P(x,y)$ for $\lambda=600$ nm are 
displayed in Figures \ref{fig:phase}  (a) and (c). Figures \ref{fig:phase} (b) and  (d) display the corresponding contour plots. Simulations are reported for fluidic lenses with circular apertures (Top row: circular aperture with horizontal  and vertical axes $a=b=1.7$ cm), 
and for fluidic lenses with elliptic apertures (Bottom row: elliptic aperture with  horizontal and vertical axes $a=1.3$ cm and $b=1.7$ cm). \\

 \subsection{Wave-front aberrations of a single membrane} 

\color{black}In order to characterize \color{black} numerically \color{black} the spectral response of optical aberrations and compare directly with experimental data, we expanded the wave-front aberrations of a single elastic membrane in terms of Zernike polynomials  up to order 14, for a beam with wavelength ($\lambda$) in the range $400-1500$ nm, \color{black} thus numerically characterizing the spectral response in the visible and infrared domains. Even though we analyze wavefront aberrations in terms of Zernike polynomials up to order 14, we only display those coefficients for Zernike polynomials which are not negligible. Namely, $P_0,P_2,P_3,P_6$ (for circular apertures) and $P_0,P_2,P_3,P_5$ (for elliptic apertures). The remaining Zernike coefficients are all below $10^{-14}$, for this reason they are not displayed in the figures. \color{black}  

\begin{figure}
\centering
\includegraphics[width=1\linewidth]{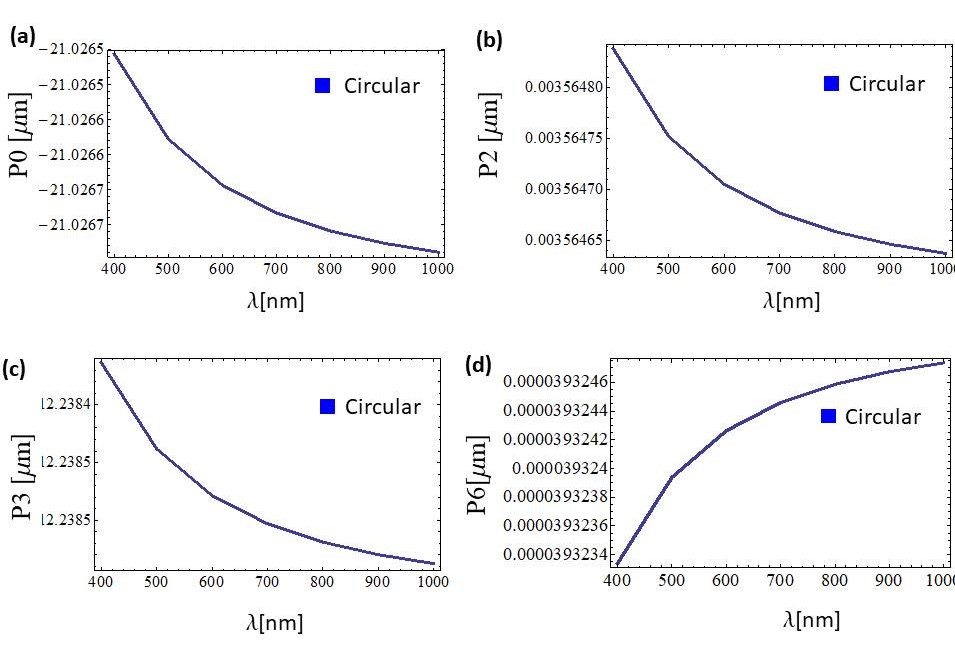}
\caption{Numerical simulations of optical aberrations based on the expansion of the wave-front in terms of Zernike polynomials over the visible range $\lambda=400-1500$ nm, for fluidic lenses with circular aperture characterized by axes ($a=1.7$ cm, $b=1.7$ cm)
 (a), (b), (c) and (d), correspond to Zernike coefficients in  $[\mu m]$ for polynomials of orders $P_0, P_2, P_3, P_6$, respectively. Zernike coefficients characterizing the remaining polynomials are negligible ($<< 10^{-14}$). Further details are in the text.} 
 \label{fig:simulationcirc}
\end{figure}

\begin{figure}
\centering
\includegraphics[width=1\linewidth]{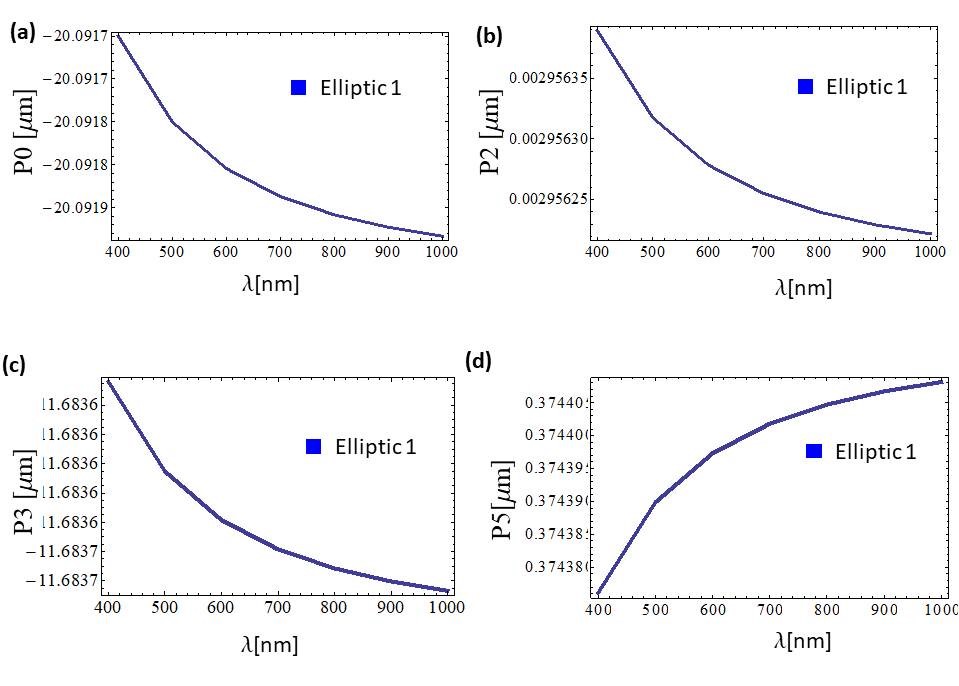}
\caption{Numerical simulations of optical aberrations based on the expansion of the wave-front in terms of Zernike polynomials over the visible range $\lambda=400-1500$ nm, for fluidic lenses with elliptic aperture characterized by axes ($a=1.5$ cm, $b=1.7$ cm)
 (a), (b), (c) and (d), correspond to Zernike coefficients in  $[\mu m]$ for polynomials of orders $P_0, P_2, P_3, P_5$, respectively. Zernike coefficients characterizing the remaining polynomials are negligible ($<< 10^{-14}$). Further details are in the text.} 
 \label{fig:simulationelli1}
\end{figure}

\begin{figure}
\centering
\includegraphics[width=1\linewidth]{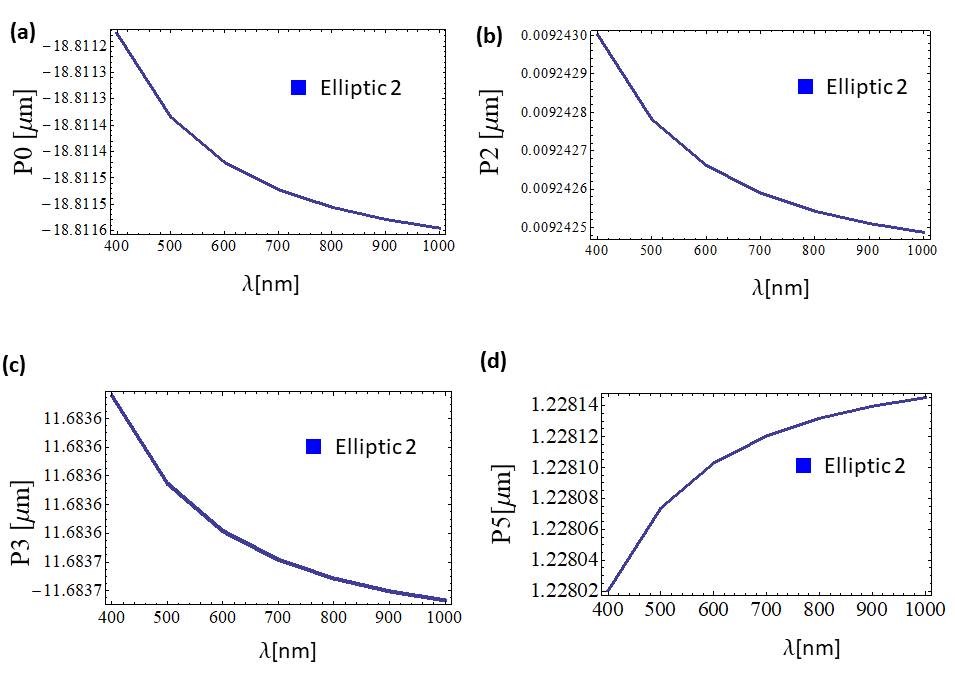}
\caption{Numerical simulations of optical aberrations based on the expansion of the wave-front in terms of Zernike polynomials over the visible range $\lambda=400-1500$ nm, for fluidic lenses with elliptic aperture characterized by axes ($a=1.3$ cm, $b=1.7$ cm)
 (a), (b), (c) and (d), correspond to Zernike coefficients in  $[\mu m]$ for polynomials of orders $P_0, P_2, P_3, P_5$, respectively. Zernike coefficients characterizing the remaining polynomials are negligible ($<< 10^{-14}$). Further details are in the text.} 
  \label{fig:simulationelli2}
\end{figure}

Numerical results of wave-front aberrations for fluidic lenses with circular aperture (axes $a=b=1.7$ cm) are displayed in Figure \ref{fig:simulationcirc}. Figures \ref{fig:simulationcirc} (a), (b), (c) and (d), correspond to normalized Zernike  coefficients for polynomials of orders $P_0, P_2, P_3, P_6$, respectively. The remaining polynomials are not reported because their coefficients are negligible ($<< 10^{-14}$). 
 Numerical results for chromatic response of fluidic lenses with elliptic apertures characterized by ellipse  axes ($a=1.5$ cm, $b=1.7$ cm) and ($a=1.3$ cm, $b=1.7$ cm) are displayed in the Figure \ref{fig:simulationelli1} and \ref{fig:simulationelli2}, respectively. 
 Figure \ref{fig:simulationelli1} and \ref{fig:simulationelli2}  (a), (b), (c) and (d), correspond to Zernike coefficients for polynomials of orders $P_0, P_2, P_3, P_5$, respectively. The remaining polynomials are not displayed because their coefficients are negligible. As it is apparent from numerical results, dependence of wave-front aberrations are of the general form $|1/\lambda|$. Moreover, spectral fluctuations in wave-front aberrations are within a fraction of $\lambda$.

\section*{Fluidic Lens Prototype}

\begin{figure}[t!]
\centering
\includegraphics[width=1\linewidth]{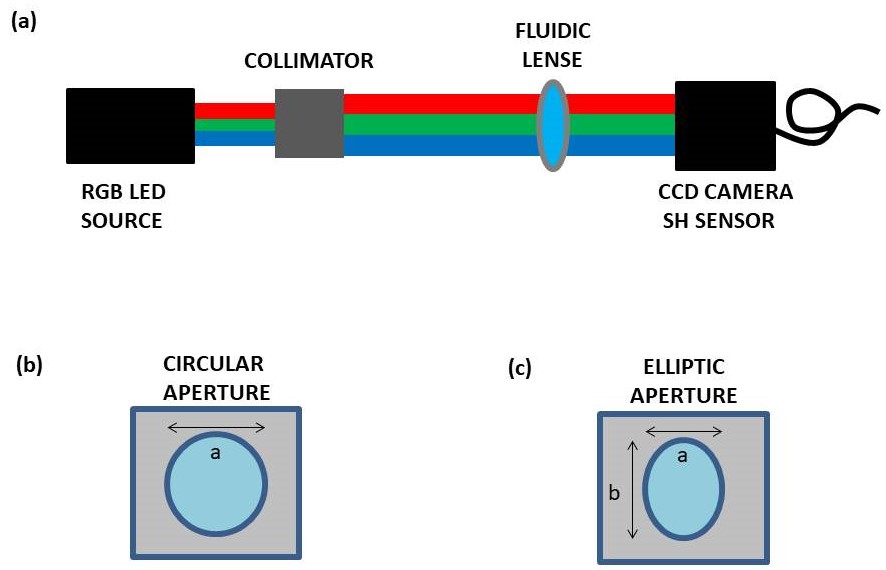}
\caption{Experimental scheme for reconstruction of the wave-front transmitted by the fluidic lens prototype, and characterization of the chromatic response of optical aberrations using a collimated incoherent programmable LED source (Alic Smart Life) and a Shack-Hartmann wave-front sensor (Thorlabs WFS150-5C). Scheme of fluidic-lense prototype: (b) circular aperture (horizontal and vertical axes $a=b=17$ mm), (c) elliptic aperture (horizontal axes ($1,2$)  $a_{1(2)}=15(13)$ mm and vertical axis $b=17$ mm). By tuning the aperture of the lens it is possible to address different optical aberrations.  } 
\label{fig:setup}
\end{figure}

As readily reported in a previous publication \cite{OSACPuentes}, the fluidic lens consists of two layers of elastic membrane of the polydimethylsiloxane (PDMS) type. 
%The fabrication is straightforward since it is possible to use a %master mold able %to repeat the procedure with precision of 0.1 %um to 10 nm fidelity \cite{Polson}. 
% By choosing a different thickness for the elastic membranes we %can manufacture %plano-convex lenses  or bi-convex lenses. 
% For the plano-convex case one of the membranes is made %significantly thicker than %the other, in order to remain %uncurved under pressure. Typical thickness for the %thick %membrane is 1200 $\mu m$, and for the thin membrane 200 $\mu m$. 
 The two elastic films are held together by an aluminum frame, sealed with the elastic membrane. An optical fluid  of refractive index matched to the polymer such as glycerol or distilled water, is injected between the elastic layers. By increasing or decreasing 
 the fluid volume mechanically injected, it is possible to tune the focal distance across several centimeters, and adjust the optical power of the lens. 
 %Due to the lightness of all the components employed the overall %weight of the prototype (without the frame) is less than 2 $gr$. 
Further, we tune one additional degree of freedom, given by the shape of the aperture. By modifying the aperture shape from circular (Fig. \ref{fig:setup} (b)) to elliptical (Fig. \ref{fig:setup} (c)), we can introduce different optical corrections. 
Typical size for the circular lens is given by a diameter $ d$=17mm, the elliptic lenses have a mayor  axis $b=17$ mm, and minor axes $a=15$ mm and $a=13$ mm. 
Further details regarding the fabrication of the fluidic lens prototype are reported in a previous article \cite{OSACPuentes}.

\begin{figure}[b!]
\centering
%\hspace{-2cm}
\includegraphics[width=1\linewidth]{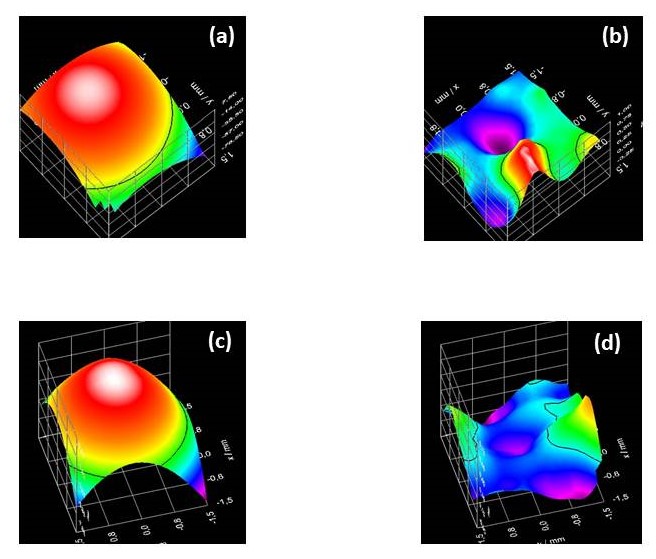}
\caption{Experimentally reconstructed wave-front using a Shack-Hartmann sensor (Model Thorlabs WFS150-5C) and a collimated incoherent LED source. (a) Reconstructed wave-front produced by a fluidic lens with circular aperture, 
(c) reconstructed wave-front produced by a fluidic lens with elliptic aperture. The qualitative difference in the wave-front due to the shape of the aperture is apparent. (b) and (d) residual difference between measured and reconstructed wave-fronts.  Further details on the Shack- Hartmann wave-front sensor are provided in \cite{OSACPuentes}. }
\label{fig:wavefront}
\end{figure}

\section*{Experimental Results}

\begin{figure}[h!]
%\hspace{-1.7cm}
\centering
\includegraphics[width=1.0\linewidth]{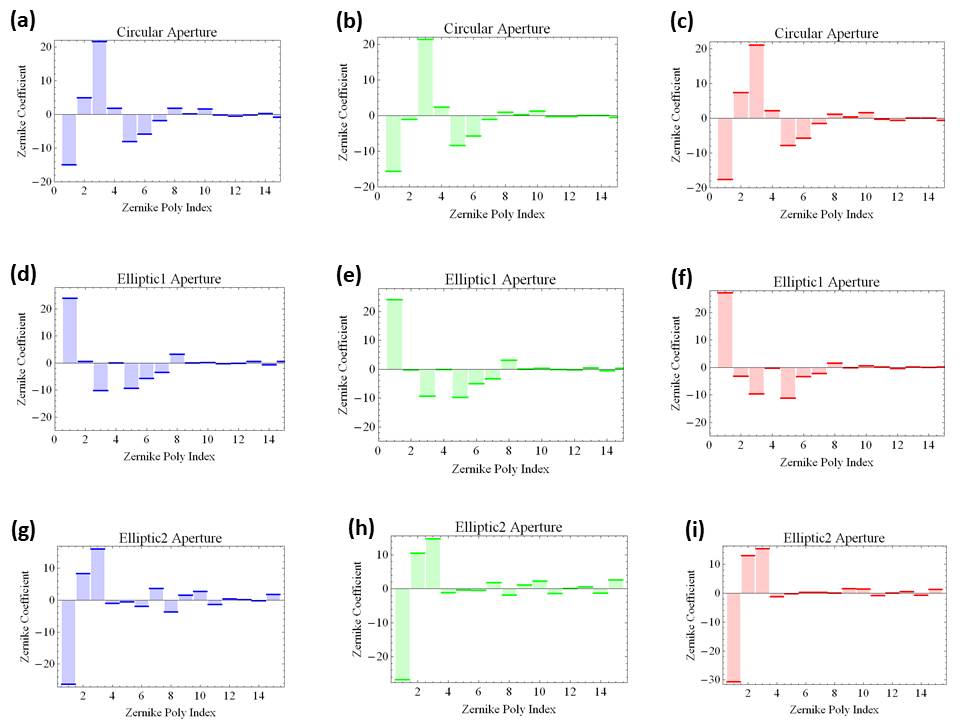}
\caption{(a) to (i) Measured aberrations in $\mu$m, in terms of coefficients of associated Zernike polynomials of order 0 to 14, for Blue, Green and Red LED light. First Column: Blue LED source, (a) Circular aperture ($a=b=1.7$ cm), (d) Elliptic aperture 1 ($a=1.5$ cm, $b=1.7$ cm), (g) Elliptic aperture 2  ($a=1.3$ cm, $b=1.7$ cm). Second Column: Green LED source, (b) Circular aperture ($a=b=1.7$ cm), (e) Elliptic aperture 1 ($a=1.5$ cm, $b=1.7$ cm), (h) Elliptic aperture 2  ($a=1.3$ cm, $b=1.7$ cm). 
Third Column: Red LED source, (c) Circular aperture ($a=b=1.7$ cm), (f) Elliptic aperture 1 ($a=1.5$ cm, $b=1.7$ cm), (i) Elliptic aperture 2  ($a=1.3$ cm, $b=1.7$ cm).The agreement with numerical simulations is mostly qualitative due to the spectral broadness of the LED source. Further details are in the text.
 }
\label{fig:zernike}
\end{figure}

\begin{figure}
\centering
\includegraphics[width=1\linewidth]{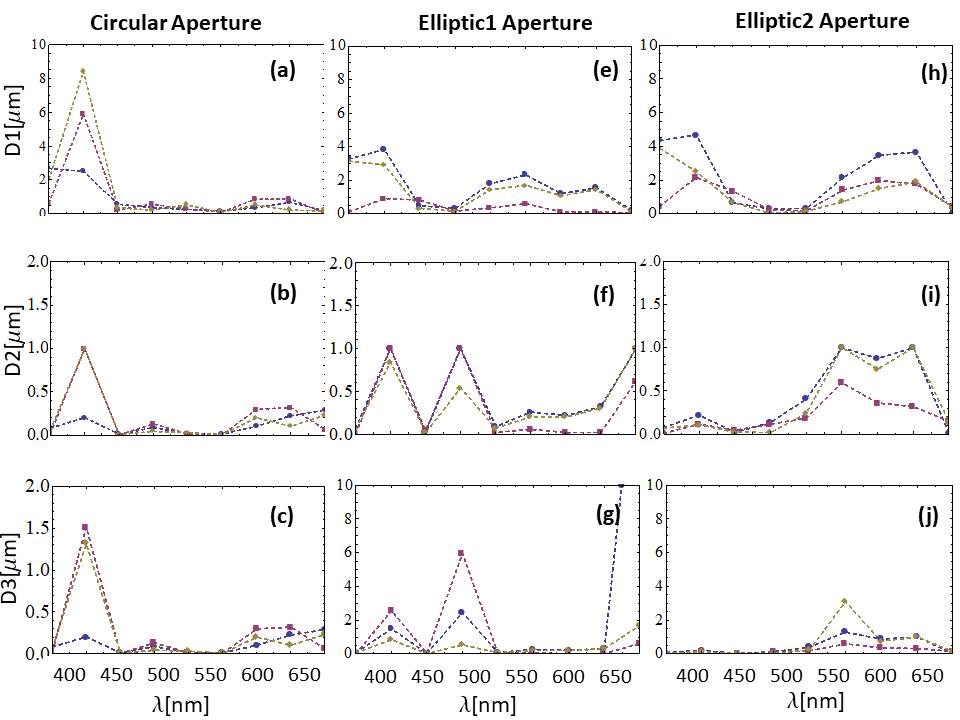}
\caption{Comparison between spectral distances between measured Zernike coefficients in $\mu$m, as quantified by three different distance measures $D_1, D_2, D_3$, for Red-Green-Blue (RGB) wavelengths. Blue markers, magenta markers, and brown markers correspond to B-R distance, B-G distance and R-G distance, respectively.  (a), (b) and (c) depict spectral distances $D_1$, $D_2$ and $D_3$ for fluidic lenses with circular apertures.  (d), (e) and (f) depict spectral distances $D_1$, $D_2$ and $D_3$ for fluidic lenses with elliptic 1 aperture,  (g), (h) and (i) depict spectral distances $D_1$, $D_2$ and $D_3$ for fluidic lenses with elliptic 2 aperture. Further details are in the text.} 
\label{fig:distance}
\end{figure}

\subsection{Wave-front reconstruction}

%Tunable optical power is a very desirable property %for any user-oriented device. In particular, for %eye-wear lenses, cameras or other machine vision %applications. One main application is, for %instance, in correcting for presbyopia. 
%Typical multi-focal crystal lenses, are static and %can only contain a limited number of focal %distances, making the field-of-view of the lens %itself very narrow.
% The tunable lens we demonstrate here can provide %for a continuum of focal distances expanding across %several centimeters without compromising the field-%of-view. Further details about the characterization %of the focusing power of the lenses are presented %in Ref. \cite{OSACPuentes}.\\
 
In order to characterize the chromatic response of the light field transmitted by the fluidic lenses, 
we reconstructed the wave-front transmitted through the lenses using a Shack-Hartmann wave-front sensor Fig. \ref{fig:setup}  (a) (Model Thorlabs WFS150-5C,  raw experimental data can be found at our Github repository \cite{github}.). To this end, we used a collimated incoherent RGB LED source (RGB: Red-Green-Blue). \color{black} Perfect collimation of a polychromatic beam can never be achieved. Here by collimation we refer to the fact that we verified the beam size did not diverge significantly over large distances (3 m or more), and we also confirmed that the wavefront impinging on the fluidic lenses was nearly a plane wavefront, so that we could use the internal calibration of the Shack-Hartmann wavefront sensor, and all measured wavefront aberrations could be ascribed to the fluidic lenses themselves. 

Shack-Hartmann wavefront sensors (SHWS) enable to analyze the shape of an incident beam's wavefront by dividing the beam into an array of discrete intensity points, using a micro-lens array. These data are then used to reconstruct and analyze the shape of the wavefront using Zernike polynomials. In addition to analyzing classical optics phenomena, they are increasingly employed in applications where real-time monitoring of the wavefront is required to control adaptive optics with the intent of removing the wavefront distortion before creating an image. In particular, SHWSs enable two types of wavefront characterizations. (I) Direct measurement (not displayed in Fig.  \ref{fig:wavefront}): shows the wavefront which is directly calculated from the measured spot deviations using a 2-dimensional integration procedure. (II) Zernike reconstruction (left column in Fig. \ref{fig:wavefront} ): displays the wavefront that is reconstructed using a selected set of the determined Zernike coefficients. The advantages of Zernike reconstruction are as follows: (i) Selecting only a few Zernike modes of lower order for reconstruction smooths the wavefront surface (noise canceling), (ii) the lowest order Zernike modes (for instance Z0 piston, Z1 tip and Z2 tilt) are always present but they are of less interest. Using an appropriate reconstruction (e.g., starting from Z3) can omit the Z0, Z1 and Z3 Zernike modes in order to see only the higher order modes. (iii) If selecting particular Zernike modes, they can be displayed and analyzed separately.  Difference (right column in Fig. \ref{fig:wavefront} ): displays the difference between the (I) directly measured wavefront and (II) reconstructed wavefront, and is therefore an indicator of the fit error.  \\

\color{black}

The incident field had a residual field curvature below $\lambda/6$. The sensor was placed 10 cm centimeters apart from the fluidic lens, with an aperture limited by the pupil size of the sensor itself, typically  3mm diameter. We reconstructed the wave-front produced by a circular fluidic lens filled with $V_{max}=6$ ml corresponding to an optical power (OP)= 50D (Fig. \ref{fig:wavefront}  (a)), and by an elliptical fluidic lens filled with $V_{min}=4$ ml corresponding to OP=36D (Fig. \ref{fig:wavefront}   (c)). The qualitative difference in the wave-front due to the shape of the aperture is apparent. Residual difference between the measured wave-front and the reconstructed wave-front are displayed in Figures \ref{fig:wavefront}  (b) and (d).  Further details on the Shack- Hartmann wave-front sensor are provided in Ref. \cite{OSACPuentes}.

\begin{figure}[h!]
%\hspace{-2cm}
\centering
\includegraphics[width=1\linewidth]{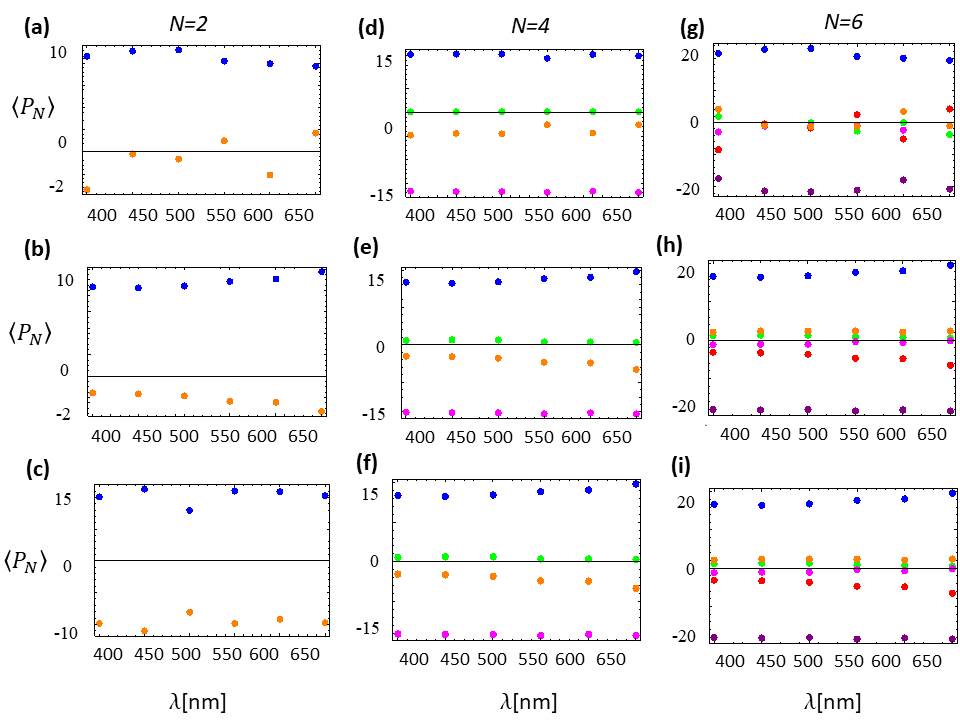}
\caption{(a) to (i) Experimental clusters for average Zernike coefficients  ($\langle P_{N}\rangle$) in the range $\lambda= 400 - 650$ nm, partitioned into a predetermined set of $N$ clusters with $N=2, 4, 6$. First Column: $N=2$ clusters. (a) Circular aperture ($a=b=1.7$ cm), (b) Elliptic aperture 1 ($a=1.5$ cm, $b=1.7$ cm), (c) Elliptic aperture 2  ($a=1.3$ cm, $b=1.7$ cm). Second Column $N=4$ clusters. (d) Circular aperture ($a=b=1.7$ cm), (e) Elliptic aperture 1 ($a=1.5$ cm, $b=1.7$ cm), (f) Elliptic aperture 2  ($a=1.3$ cm, $b=1.7$ cm). Third Column: $N=6$ clusters. (g) Circular aperture ($a=b=1.7$ cm), (h) Elliptic aperture 1 ($a=1.5$ cm, $b=1.7$ cm), (i) Elliptic aperture 2  ($a=1.3$ cm, $b=1.7$ cm). Further details are in the text.
}
\label{fig:clusters}
\end{figure}

\subsection{Measured Zernike Coefficients}

In order to experimentally characterize the spectral response of optical aberrations in the central region of the fluidic lens prototype, we use the experimental setup described in Fig. \ref{fig:setup}  (a). A collimated incoherent beam, produced by a programmable LED source (Alic Smart Life, 14W, Luminous Flux 1400 lm, $\lambda=400-1045$ nm) propagates through the fluidic lens and is imaged by a Shack-Hartmann wave-front sensor (Model Thorlabs WFS150-5C), located at a distance of 2 cm from the fluidic lens, in order to image the near field produced by the lens. The area of the beam to be characterized is determined by the aperture of the sensor (typically 3mm). We verified that the transverse profile of the beam did not change significantly when tuning the wavelength of the source across the entire spectral range. Spectral characterizations in the visible range are mostly qualitative due to the broad spectrum produced by the incoherent LED source. \\

Measured aberrations in $\mu$m, in terms of the coefficients associated with Zernike polynomials of order 0 to 14, for Red, Green and Blue LED illumination are displayed in Figures \ref{fig:zernike}   (a) to (i). First Column: Blue LED source, (a) Circular aperture ($a=b=1.7$ cm), (d) Elliptic aperture 1 ($a=1.5$ cm, $b=1.7$ cm), (g) Elliptic aperture 2  ($a=1.3$ cm, $b=1.7$ cm). Second Column: Green LED source, (b) Circular aperture ($a=b=1.7$ cm), (e) Elliptic aperture 1 ($a=1.5$ cm, $b=1.7$ cm), (h) Elliptic aperture 2  ($a=1.3$ cm, $b=1.7$ cm). 
Third Column: Red LED source, (c) Circular aperture ($a=b=1.7$ cm), (f) Elliptic aperture 1 ($a=1.5$ cm, $b=1.7$ cm), (i) Elliptic aperture 2  ($a=1.3$ cm, $b=1.7$ cm). \\

\begin{figure}[h!]
\centering
%\hspace{-1.5cm}
\includegraphics[width=1\linewidth]{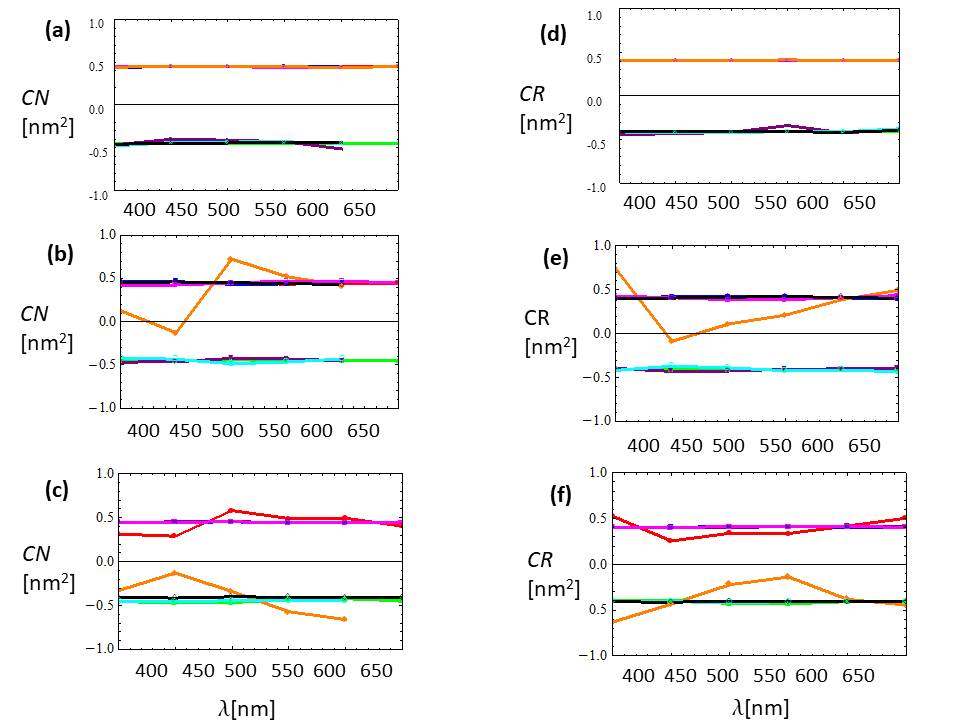}
\caption{(a) to (f) Normalized convolution ($CN$) and correlation ($CR$) between wavelength ($\lambda$) and measured Zernike coefficients ($P_{s}$). Left column: $CN[nm^2]$. (a) Circular aperture ($a=b=1.7$ cm), (b) Elliptic aperture 1 ($a=1.5$ cm, $b=1.7$ cm), (c) Elliptic aperture 2 ($a=1.3$ cm, $b=1.7$ cm). Rigth column: $CR[nm^2]$. (a) Circular aperture ($a=b=1.7$ cm), (b) Elliptic aperture 1 ($a=1.5$ cm, $b=1.7$ cm), (c) Elliptic aperture 2 ($a=1.3$ cm, $b=1.7$ cm). Further details are in the text.}
\label{fig:cncr}
\end{figure}

\color{black}In order to quantify the agreement between experimental results and numerical simulations, we calculated the distance between measured Zernike coefficients for Red-Green-Blue (RGB) wavelengths. We considered three different distance measures defined for two sets of data $\{a,b,c\}$ and $\{x,y,z\}$ in the form: (1) Euclidean distance $D_1=\sqrt{|a - x|^2 + |b - y|^2 + |c - z|^2}$, (2) Canberra distance $D_2=|a - x|/(|a| +|x|) + |b - y|/(|b| + |y|) +  |c - z|/(|c| + |z|)$, and (3) Bray-Curtis distance $D_3=(|a - x| + |b - y| + |c - z|)/((|a + x| + |b + y| + |c + z|)$. 

\color{black}
A brief comparison between the different distance measures is in order: $D_1$ corresponds to the “Pitagoric” distance, and is the only measure which can be subject to a direct geometrical interpretation, therefore in this sense it is the most intuitive one. $D_2$ and $D_3$ distance measures are similar in essence, they are both based on the algebraic concept of norm of a vector. They differ in the normalization factor, while $D_2$ normalizes each element of the vector independently, $D_3$ introduces a global normalization factor, and is therefore less sensitive (larger in modulus), as it can be verified in Figure \ref{fig:distance} (g) and (j). Note that $D_1$ is not normalized, for this reason it is typically larger in modulus than $D_2$ and $D_3$. We did not include the Manhattan distance in this analysis because it returned practically identical results to the Euclidean distance. The usefulness of the Manhattan measure was clearly revealed when employed in clustering techniques (see Section 2.4, Figure  \ref{fig:clusters}). 
\color{black}

Comparison between spectral distances for measured Zernike coefficients in $\mu$m, for RGB wavelenghts are displayed Figure 10. Blue markers, magenta markers, and brown markers correspond to B-R distance, B-G distance and R-G distance, respectively. Figure 10 (a), (b) and (c) depict spectral distances $D_1$, $D_2$ and $D_3$ for fluidic lenses with circular apertures. Figures 10 (d), (e) and (f) depict spectral distances $D_1$, $D_2$ and $D_3$ for fluidic lenses with elliptic1 aperture, and Figs. 10  (g), (h) and (i) depict spectral distances $D_1$, $D_2$ and $D_3$ for fluidic lenses with elliptic2 aperture.  Distances are within a fraction of the wavelength, in agreement with numerical simulations.

\subsection{Partition into Clusters}

In order to further classify the spectral response of optical aberrations, we partitioned the data into a predetermined number of clusters ($N$) across the visible range ($\lambda=400-650$ nm). More specific, Zernike coefficients were partitioned into subgroups (or clusters) representing proximate collections of elements based on a distance or dissimilarity function. In particular, we consider the Manhattan distance, given by the sum of the absolute difference between the elements. Identical element pairs have zero distance or dissimilarity and are grouped into a given cluster,  all others have positive distance or dissimilarity.

\color{black}
Clustering techniques provide for a robust quantitative tool to classify large sets of data according to the distance between the elements in the clusters. This, in turn, can enable to identify emerging trends in experimental data. In addition, they can enable quantitative comparisons between experiment and numerical/theoretical predictions. Moreover, in our case, we performed clustering techniques based on an alternative distance measure, e.g., the Manhattan distance, which provided for further insights into the way in which averaged experimental data are grouped and distributed, according to the input wavelength. For instance, from the clustering analysis one can infer that average positive and negative Zernike coefficient are typically distributed with similar probabilities, for all input wavelengths. Note that the insights provided by clustering techniques are complementary to the direct calculations of distances between elements (Figure  \ref{fig:distance}).

\color{black}
Experimental clusters for average Zernike coefficients  ($\langle P_{N} \rangle$) in the range $\lambda= 400 - 650$ nm, partitioned into a predetermined set of $N=2, 4,$ and $ 6$ clusters are presented in Figures  \ref{fig:clusters} (a) to (i). First Column: $N=2$ clusters. (a) Circular aperture ($a=b=1.7$ cm), (b) Elliptic aperture 1 ($a=1.5$ cm, $b=1.7$ cm), (c) Elliptic aperture 2 ($a=1.3$ cm, $b=1.7$ cm). Second Column $N=4$ clusters. (d) Circular aperture ($a=b=1.7$ cm), (e) Elliptic aperture 1 ($a=1.5$ cm, $b=1.7$ cm), (f) Elliptic aperture 2  ($a=1.3$ cm, $b=1.7$ cm). Third Column: $N=6$ clusters. (g) Circular aperture ($a=b=1.7$ cm), (h) Elliptic aperture 1 ($a=1.5$ cm, $b=1.7$ cm), (i) Elliptic aperture 2  ($a=1.3$ cm, $b=1.7$ cm). For $N=2$, Zernike coefficients can be classified into two main clusters, corresponding to either an average  positive amplitude $\langle P_{2} \rangle=+10$ (blue dots), or an average  negative amplitude $\langle P_{2} \rangle=-2$ (orange dots). Next, for $N=4$, 
Zernike coefficients can be classified into 4 clusters, one with an average positive amplitude $\langle P_{4} \rangle=+15$ (blue dots), one with an average negative amplitude $\langle P_{4} \rangle=-15$ (magenta dots), and remaining two with nearly vanishing amplitudes $\langle P_{4} \rangle \approx 0$ (green and orange dots). Finally, for $N=6$, Zernike coefficients are classified into 6 clusters, one with an average positive amplitude $\langle P_{6} \rangle=+20$ (blue dots), one with an average negative amplitude $\langle P_{6} \rangle=-20$ (purple dots), and the remaining 4 clusters with nearly vanishing amplitudes $\langle P_{6} \rangle \approx  0$ (green, orange, magenta and red dots). The decreasing amplitude of the average Zernike coefficients for decreasing the number of clusters ($N$) can be ascribed to averaging over a broader range of amplitudes, since reducing $N$ increases the diversity of the elements.

\subsection{Convolution and Correlation}

\color{black}
Convolution (CN) and correlation (CR) measurements are robust analytical tools which enable quantitative analyses of the interrelation between two experimental magnitudes. In this case, Zernike coefficients vs. wavelength. Specifically, CR/CN=+1(-1) represents a maximal positive(negative) interrelation, while CR/CN=0 represent no interrelation at all. Moreover, these methods enable to identify emerging trends, or salient features for specific values of the measured quantities. In addition, they enable direct contrast and comparison with other characterization methods, such as clustering techniques, and with theoretical/numerical predictions. From the CR and CN data, we can conclude that the correlation between measured Zernike coefficients and wavelength is typically medium CR/CN=+0.5(-0.5), uniformly distributed between positive and negative values for all wavelength, with no specific wavelength-dependent salient features. These results are in agreement with the conclusions obtained from clustering techniques, and from numerical predictions.

\color{black}
In both CN and CR the basic idea is to combine a kernel list with successive sub-lists of a list of data. The convolution of a kernel $K_{r}$ with a list $u_{s}$ has the general form $\sum_{r} K_{r} u_{s-r}$, while the correlation has the general form  $\sum_{r} K_{r} u_{s+r}$. In particular, for a kernel list $K_{r}=[x,y]$ and list of data $u_{s}=[a,b,c,d,e]$, the convolution ($CN$) results in the combined list:

\begin{equation}
CN=[b x + a y, c x + b y, d x + c y, e x + d y ],
\end{equation}

while the correlation ($CR$) results in the combined list:

\begin{equation}
CR=[a x + b y, b x + c y, c x + d y, d x + e y].
\end{equation}

We calculated the convolution ($CN$) and correlation ($CR$) between the wavelength ($\lambda$) and the measured Zernike coefficients ($P_{s}$), where $s=0,...,14$ labels the polynomial order in each sub-list. We consider a kernel specified by the wavelength range $K_{r}=[400,450,500,550,600,650]$ in nm, and list of measured Zernike coefficients ($P_{s}(\lambda)$) in nm, for each different input wavelength ($\lambda$), of the form  $u_{s}=[P_{s}(400), P_{s}(450), P_{s}(500), P_{s}(550), P_{s}(600), P_{s}(650)]$. A plot of the normalized correlation ($CR$) and convolution ($CN$) are presented in Figures \ref{fig:cncr} (a) to (f). Left column: $CN[nm^2]$. (a) Circular aperture ($a=b=1.7$ cm), (b) Elliptic aperture 1 ($a=1.5$ cm, $b=1.7$ cm), (c) Elliptic aperture 2 ($a=1.3$ cm, $b=1.7$ cm). Right column: $CR[nm^2]$. (d) Circular aperture ($a=b=1.7$ cm), (e) Elliptic aperture 1 ($a=1.5$ cm, $b=1.7$ cm), (f) Elliptic aperture 2 ($a=1.3$ cm, $b=1.7$ cm). As a general trend, Zernike polynomials display either a positive correlation with $\lambda$ ($CR/CN=+0.5$), or a negative correlation with $\lambda$ ($CR/CN=-0.5$). Fluctuations on this trend increases as the asymmetry in the ellipse axes increases. The significant color spread indicates that there is no particular correlation, neither positive nor negative, between the Zernike order ($s$) and wavelength ($\lambda$).

\section*{Discussion}

We have presented a comprehensive numerical and experimental study of the spectral response of optical aberrations in macroscopic fluidic lenses with high dioptric power, tunable focal distance, and aperture shape \cite{OSACPuentes}. Our investigation is based on an empirical characterization of the optical and material properties of thin elastic membranes, in particular of the refractive index of polymers, such as PDMS, according to the first order Sellmeier model \cite{Schneider}. Using a Shack-Hartmann wave-front sensor we experimentally 
reconstructed the near-field wave-front transmitted by such fluidic lenses, and we characterized the chromatic response of optical aberrations in terms of Zernike polynomials over the visible wavelength range ($\lambda=400-650 $ nm), using an incoherent programmable LED source. 
Moreover, we further classified the spectral response of the lenses using clustering techniques, encountering that for a pre-determined number of clusters ($N=2,4,6$), the Zernike coefficients characterizing the spectral response can be classified in three main clusters over the entire wavelength range. Namely, a cluster with positive Zernike coefficients, a cluster with negative Zernike  coefficients, and a cluster with nearly vanishing Zernike coefficients.
 In addition we performed correlation ($CR$) and convolution ($CN$) measurements, finding that as a general trend Zernike polynomials display either a positive correlation with $\lambda$ ($CR/CN=+0.5$), or a negative correlation with $\lambda$ ($CR/CN=-0.5$). Fluctuations on this trend increases as the asymmetry in the ellipse axes increases. Experimental results are in agreement with our theoretical model of the non-linear elastic membrane deformation. A complete characterization of
 the spectral response of optical aberrations for coherent illumination will be presented in an upcoming work.

\section*{Acknowledgements}
The authors are grateful to the Solar Energy Department (TANDAR-CNEA)  and to the Laboratory of Polymers (FCEN-UBA) for assistance in PDMS membrane preparation. GP gratefully acknowledges financial support from
 PICT2014-1543, PICT2015-0710 Startup, UBACyT PDE 2016, UBACyT PDE 2017.

\end{document}